\documentstyle[11pt,aaspp4]{article}

\slugcomment{Astrophysical Journal Letters, in press}
\lefthead{Montoya et al.}
\righthead{Lensing in Compact Groups}

\begin{document}

\title{The Surface Density Profiles and Lensing Characteristics
of Hickson's Compact Groups of Galaxies}

\author{M. L. Montoya and R. Dom\'{\i}nguez-Tenreiro} 
\affil{Dpt. F\'{\i}sica Te\'orica C-XI, 
Universidad Aut\'onoma de Madrid, \\
E-28049 Cantoblanco, 
Madrid, 
Spain}

\author{G. Gonz\'alez-Casado}
\affil{Departament de Matem\`atica Aplicada II,
Universitat Polit\`ecnica de Catalunya, \\ 
Pau Gargallo 5,
E-08028, 
Barcelona, 
Spain}

\author{G. A. Mamon}
\affil{Institut d'Astrophysique (CNRS),
98 bis Blvd. Arago, F-75014,
Paris, 
France; \\
also DAEC,
Observatoire de Paris-Meudon, 
F-92195 Meudon, 
France} 

\and 

\author{E. Salvador-Sol\'e}
\affil{Departament  d'Astronomia i Meteorologia,
Universitat de Barcelona, 
Avda. Diagonal 647,
E-08028 Barcelona, 
Spain}

\begin{abstract}
A statistical method is developed to infer the typical density profiles of poor
galaxy systems without resort to binning of data or assuming a given center
to each system.
The method is applied to the accordant redshift quartets in Hickson's
compact groups (HCGs).
The distribution of separations in these groups is consistent
with a unique generalized
modified Hubble surface density profile, with best-fit
asymptotic slope
$\beta = 1.4$ and core radius $R_c = 18 \, h^{-1} \, \rm kpc$, although
a King profile ($\beta = 1$) is also consistent with the data (with $R_c = 6
\, h^{-1} \, \rm kpc$).
These distributions are more concentrated than what has been previously
determined for these groups. 
HCGs are unlikely to act as strong gravitational lenses, but analogous
systems 5 to 10 times more distant should produce a non-negligible fraction
of giant arcs.
\end{abstract}


\section{INTRODUCTION}

The 100 compact groups (CGs)
cataloged by Hickson (1982) have been among the most
systematically observed galaxy systems, and yet, there is little consensus on
the nature of these systems, between those who believe that most of Hickson's
compact groups (HCGs) are roughly as dense in 3D as they appear to be in
projection (e.g. Hickson \& Rood 1988), and those who argue that
CGs are mostly chance alignments of galaxies along the line of
sight, within larger `loose' groups (Mamon 1986) or cosmological filaments
(Hernquist, Katz \& Weinberg 1995).

The nature of CGs may be clarified through 
the characterization of their galaxy surface number density profiles,
$\Sigma(R)$. 
Indeed, if CGs are caused by chance alignments within much
larger systems, then the galaxy positions should be relatively random, and
the corresponding density profiles should be roughly homogeneous.
The galaxy number density profile in CGs
is also interesting by itself, as compared with the density profiles
of the gaseous and dark matter components, in connection with the theory of 
the formation and evolution of galaxies within groups.

The small number of galaxies (typically 4) per HCG makes it very difficult to
estimate the density profile.
Hickson et al. (1984) have shown that all 100 HCGs, rescaled to
the same size, show a concentrated density profile.
More recently, Mendes de Oliveira \& Giraud (1994, MG94) have 
argued that HCGs, once rescaled to a common size, are consistent with
a King (1962) surface density profile, 
with typical core radius of $15 \, h^{-1} \, \rm kpc$.
Unfortunately, such studies suffer from the uncertainty introduced by
the definition of a center for a
system of typically four galaxies, and from the
assumption that this center is
at the galaxy barycenter.
Nevertheless, there are statistical methods to determine the density profiles
of discrete systems that do not make use of the definition of the system
center, relying instead upon the distribution of projected particle (galaxy)
separations 
(see the mathematical foundations in
Salvador-Sol\'e, Sanrom\`a $\&$ Gonz\'alez-Casado  1993, SSG;
Salvador-Sol\'e, Gonz\'alez-Casado $\&$ Solanes 1993, SGS).

Alternatively, one can obtain good estimates of the surface density
profiles of the X-ray emitting diffuse intergalactic gas, which, so far,
has been firmly
detected with the ROSAT PSPC in 7
HCGs out of 17 pointings
(see Mulchaey et al. 1996 [MDMB], and references therein),
suggesting that those HCGs
are physical bound entities.
These ROSAT
observations reveal
projected
gas density profiles consistent with 
$\Sigma_{\rm gas}(R) = \Sigma_{\rm gas}^0 / [1 + (R/R_c)^2]^{(3\beta_{\rm
gas}-1)/2}$, 
and with core radius
$4 \leq R_c \leq 30 \, h^{-1} \, \rm kpc$ and $0.38 \leq \beta_{\rm gas} \leq
0.92$ 
(see MDMB, and references therein).
In comparison, Abell clusters have X-ray core radii typically 15 times larger
(Jones \& Forman 1984; Mushotzky 1994).

The efficiency of a structure as a gravitational lens is determined by its
projected ${\it mass}$ density profile, $\Sigma_{\rm mass}(R)$, and the
relative
lens-observer and source-observer distances (Bourassa \& Kantowski 1975).  
Extrapolating the small intergalactic gas core radii to the distribution of
total binding mass suggests that CGs may be concentrated enough to 
be candidates for
gravitational lenses (MG94). 
If the binding mass in groups is more concentrated than the intergalactic
gas, as is the case in clusters
(Henriksen 1985; Hughes 1989; Durret et al. 1994),
the lensing efficiencies of groups are enhanced.

Hence, various 
signatures of 
gravitational lensing (giant arcs, arclets, shear,
apparent magnitude enhancement) in CGs of existing catalogs
(Hickson et al. 1992; Prandoni, Iovino $\&$ MacGillivray 1994) 
or of future deeper surveys, if observed,
would be of major interest because 
 they would allow a
quantification of the dark matter distribution in CGs through
a method independent of X-ray analyses.
This quantification has important consequences for the theories of formation
and evolution of galaxies and their groupings in the framework of the
large scale structure of the Universe.

In this Letter, we implement a statistical method to determine average galaxy
density profiles in the case of poor statistics,
with no assumed
group center.
The mean galaxy surface number density profile of compact
groups is computed from these
statistical methods, 
 and the implications for the lensing properties of compact
groups are studied.

\section{THE GALAXY SURFACE DENSITY PROFILE}

For a catalog of groups (whose centers are uncorrelated),
the galaxy two-point  correlation function, $\xi(s)$, is
(Peebles 1980):
\begin{equation}
\xi(s) = {{ \langle N_{i}^{2} (\Sigma_{i}*\Sigma_{i}) (s)
\rangle n_{G}}\over{n^{2}}}\ ,
\label{eq:xi}
\end{equation}
where
$\Sigma_{i}(s)$ is the projected galaxy density profile of the
{\it i$^{\underline{th}}$\/} group normalized to unity,
$s$ is the projected intergalactic separation,
$N_{i}$ is the number of  galaxies in the  {\it i$^{\underline{th}}$\/}
group,
$n_{G}$ and $n$ are the mean number density of groups and galaxies
respectively,
$\langle \,\,\rangle$  is the average taken  over the
groups of the catalog and
$(\Sigma_{i} * \Sigma_{i})(s)$ is the convolution of the projected
density profile of the  {\it i$^{\underline{th}}$\/} group,
which can be calculated by means of the deconvolution method
(SSG; SGS):
\begin{equation}
(\Sigma * \Sigma)(s)= {\cal F}_{1} \circ {\cal A} \left\{ {\cal A}
\circ  {\cal F}^{-1}_{1}
\left[2P(\ge s)\right]\right\} \ ,
\label{eq:conv}
\end{equation}
where ${\cal F}_{1}$ and ${\cal A}$ are the 1D Fourier
and Abel
transforms, respectively,
and $P(\ge s)$ is the 
fraction of galaxy pairs with
intergalactic separation $\geq s$.

The average galaxy surface density profile for 
groups in a given catalog can be determined 
through a fitting procedure by means of equations
(\ref{eq:xi}) and (\ref{eq:conv})
above. 
Assume that the groups
in the catalog   
are discrete realizations of an analytical profile, 
$\Sigma(R)$. 
We use here the
{\it generalized modified Hubble profiles} (hereafter, GMHP,
see Jones \& Forman 1984), 
whose shape is characterized by 
the core radius, $R_c$, and the slope, $\beta$:
\begin{equation}
\Sigma(R;R_c,\beta) = {\Sigma_{0}\over \left[ 1 + \left(R/R_c\right)^{2}
\right]^{(3\beta - 1)/2}}\ ,
\end{equation}
where $\Sigma_{0}$ is the central projected density. 
Our free parameters $R_c$ and $\beta$ are obtained by calculating
$\xi^{\rm data}(s)$ for the 
data catalog and $\xi^{T}(s;R_c,\beta)$ for the analytical profile,
using equations (\ref{eq:xi})
and (\ref{eq:conv}), 
and minimizing
\begin{equation}
X^{2}(R_c,\beta) = {1\over N_{\rm sep}}
\sum_{k=1}^{N_{\rm sep}} 
{[\xi^{\rm data}(s_k)-\xi^{T}(s_k;R_c,\beta)]^2\over \sigma_k^{2}}\ ,
\label{eq:X2}
\end{equation}
where $N_{\rm sep}$ is the number of bins of intergalactic separations, $s_k$.
The dispersions
$\sigma_k^{2} = \langle (\xi_{k}-\langle \xi_{k} \rangle)^{2}\rangle$,
required in evaluating equation (\ref{eq:X2}),
are evaluated from 
Monte-Carlo simulations of
artificial group catalogs,
generated with GMHPs,
with the same number of galaxies per group 
and of groups
per catalog 
as the real group catalog. 
Since
the different $\xi_k$ 
are
not independent random variables, the number of
degrees of freedom of $X^2$ is not known, and we have to
resort to Monte-Carlo simulations to evaluate empirically its  
distribution (instead of the usual $\chi^2$ statistics), 
for estimating confidence levels.
Note that only
information on the projected intergalactic separations is required in
the deconvolution method. This circumvents the delicate problem
of determining the center of the groups. Moreover, noise suppression
by filtering is also implicit in this method.

In catalogs of groups, one has a low number of galaxies per   
group and of groups per catalog.  
In this case of {\it poor statistics}, the  reliability of the method has
been
tested with
simulations of artificial catalogs whose groups are realizations
of 
GMHPs with fixed $\beta$.
Our results show that the method is 
 able to recover the input value 
of $R_c$ to a typical
90\% confidence precision of a factor 1.6, 
and 
 to distinguish between different 
values of $\beta$
in the case of 
poor statistics (Montoya et al. 1995).

The method explained  
above has
been applied to the HCG sample of 42 accordant redshift quartets
(Hickson et al. 1992). 
We have tested the null hypothesis ${\cal H}_0$ that the groups 
of this sample are realizations of a unique 
GMHP with $1 \le \beta \le 3$. 
For each value of $\beta$, we have calculated
$X^2$ (eq. [\ref{eq:X2}]) for the sample, and the cumulative
probability $F(X^2)$ from
Monte-Carlo simulations. The resulting
confidence levels in the 
($R_c$, $\beta$) plane 
are plotted in Figure 1.

The optimum fit is obtained for 
$\beta = 1.4$ and $R_c = 18 \,h^{-1}$ 
kpc (hereafter, model 2), corresponding to $F(X^2) = 0.16$. 
For $\beta = 1, 2,$ and 3, the best fit is for 
$R_c = 5.7, 32.7,$ and $50.1\,h^{-1}\,\rm kpc$ (models 1, 3, and 4
respectively). 
Our fitted surface density profiles have been
obtained without any rescaling of the sizes of the groups. Such
rescalings are not necessary because we get good fits 
without them. 
To 
show this,
we have compared
the distributions of the minimum, median and maximum intergalactic 
separations for each group in the catalog, with those 
for artificial 
catalogs of 42 quartets, each of them being a realization of a unique 
GMHP with $\beta = 1.4$ and $R_c = 18\, h^{-1}$ kpc.
Table 1 shows the median Kolmogorov-Smirnov
probabilities of not being able to reject the
hypothesis that the distribution of separations of the catalog are consistent
with that of the Monte-Carlo catalogs with $\beta = 1.4$ and $R_c = 18\,
h^{-1}$ kpc.
We conclude that any rescaling (using the minimum, median and maximum
intergalactic separations 
as typical lengths) leads, indeed, to much poorer fits, whatever values of
both  
$\beta$ and $R_c$.

MG94 have also computed the galaxy surface number density profile 
for a sample of HCGs.
Their fitting procedure requires a group center (for which they adopt the
barycenter), and, moreover, they scale all groups in their sample to
a common size by normalizing each group
to the largest galaxy distance to the group barycenter,
$R_{\rm max}$.
Both
of these operations introduce important uncertainties in their results.
On one hand,
the center of a system of few (typically 4) galaxies 
is not well defined.
Beers and Tonry (1986) have shown that any
possible inner galaxy concentration is washed out when 
the barycenter is used as the center of galaxy clusters.
So, given $\beta = 1$, MG94's core radius should be much larger than ours,
and indeed it is 2.5 times larger
(Fig. 1).
On the other hand, 
it is not clear that groups should have a common size and that $R_{\rm max}$
should be a useful measure of that size.
Instead, our results show that our sample of HCGs can be considered
as
a realization of a unique profile,
and that the dispersion in $R_{\rm max}$ arises from statistical
fluctuations.

The galaxy density profiles found here are steeper than 
the gas density profiles obtained from X-ray surface brightness 
data (see \S\ 1).
Qualitatively, this 
can be expected in the framework of the $\beta$-model 
(Cavaliere 1974), which predicts
$\beta_{\rm spec} = \sigma_v^2/(kT/\mu m_p)
= \beta_{\rm gas} / \beta_{\rm gal}$.
Indeed, the galaxy system is cooler than the X-ray emitting gas, as 
HCGs have $\beta_{\rm spec} \simeq 0.7$ (Mamon \& Henriksen 1996).
The typical values of $\beta_{\rm gas} =
0.6$ and our best fit of $\beta_{\rm gal} = 1.4$ are
consistent with $\beta_{\rm spec}$ in the context of the $\beta$-model, given
the uncertainties in these three parameters.
Hydrodynamical 
simulations of X-ray cluster formation including galaxy formation 
(e.g. Frenk et al. 1996) 
also predict a velocity bias and spatial gas/galaxy segregation.

\section{LENSING BY COMPACT GROUPS}
Due to their peaked density distributions (\S\ 2), HCGs
could
constitute efficient gravitational lenses.
An HCG at angular distance $D_l$
with central projected mass density $\Sigma_{\rm mass}(0)$
may produce an elongated arc image 
of background sources at a 
distance $D_{s}$ 
if (Bourassa \& Kantowski 1975) $\Sigma_{\rm mass}(0) \ge \left(c^2/4\pi G
\right) \left(D_s/\left[D_l D_{ls} \right]\right)$ 
(strong lensing condition, hereafter SLC).
For $\Omega= 1$ and a smooth universe, the angular distance is $D_i = 2(c/H_0)
(G_i-1)/G_i^3$, where $G_i = (1+z_i)^{1/2}$, $i=l,s$, and one can write
(Blandford \& Kochanek 1987)
$D_{ls} = 2(c/H_0) (G_s-G_l)/(G_s^3 G_l)$.
The SLC imposes, for a given lens, 
a minimum source redshift:
\begin{equation}
z_{s}^{\rm min} =  \left [{A G_l (G_l-1)-G_l^4 \over A (G_l-1)-
G_l^4}\right]^2-1 \ , 
\label{eq:zsmin}
\end{equation}
where $A = 8\pi G \Sigma_{\rm mass}(0)/ (c H_0)$. 
A physical solution $ z_{s}^{\rm min} > z_{l}$ is obtained, after some
algebra, directly from the SLC when  
$ \Sigma_{\rm mass}(0) >   (1+z_l)^{1/2}  \left[c^2/\left(  4\pi G
D_l \right)\right]$.
Assuming that the galaxy system has isothermal and isotropic kinematics, one
can solve for the total mass density in Jeans equation, and
integrating over the on-axis line-of-sight, one finds
$\Sigma_{\rm mass}(0) = 3\beta\sigma_v^2  / (2 G R_c)$,
where $\sigma_v$ is the group 1D velocity dispersion.
Dropping the factor $(1+z_l)^{1/2} \simeq 1+z_l/2$,
the  condition for $ z_{s}^{\rm min} > z_{l}$ can now be written:
\begin{equation}
D_l \,\left ({\sigma_v \over c}\right)^2
> {R_c \over 6\pi\,\beta}  \ .
\label{eq:dlstrong}
\end{equation}

In Figure 2, we plot ({\it thick\/} histograms)
the cumulative fraction of HCGs\footnote{Correcting Hickson et al.'s velocity
dispersions by $[n/(n-1)]^{1/2}$} from our sample 
for which the
product $D_l (\sigma_v/c)^2$ exceeds different thresholds
(eq. [\ref{eq:dlstrong}]),  marked by {\it arrows\/}.
For model 1, three HCGs (8, 60, and 82) in our sample verify equation
(\ref{eq:dlstrong}), 
implying (eq. [\ref{eq:zsmin}])
$z_s > 0.19, 0.20$ and 0.08,
respectively. 
For our other models, no HCG in our sample fulfills the 
strong lensing condition.
The three HCGs that are potential strong gravitational lenses
have much higher line-of-sight
velocity dispersion than the 
average (517, 472, and $714\,\rm km \, s^{-1}$, compared to a median of
$237 \, \rm km \, s^{-1}$ for our HCG sample), 
and this could easily
be due to sampling and projection effects, 
as is suggested by the lack of HCGs detected in
X-rays with gas hotter than $1\,\rm keV$ (MDMB).
Put another way, 
since most HCGs detected in X rays have temperatures close to $0.9\,\rm keV$
(see MDMB), with Mamon \&
Henriksen's (1996) estimate of $\beta_{\rm spec}$, we infer {\it true\/} 1D
velocity dispersions of $\sigma_v = 315 \, \rm km \, s^{-1}$.
The {\it thin\/} histograms in Figure 2 illustrate the situation
if {\it all\/} HCGs have $\sigma_v = 315 \, \rm km \, s^{-1}$,
and we infer that no HCGs induce strong gravitational lensing.

HCGs are relatively nearby ($z_{\rm median} = 0.03$)
 and future catalogs should provide more distant
CGs.
In light of this, we have considered a sample of CGs with the same
properties as before, but with $D_l \sigma_v^{2}$ multiplied
by a factor of 5 and 10, and we have plotted in Figure 2 the corresponding 
cumulative fractions of CGs whose $D_l (\sigma_v/c)^2$ values
exceed  the same thresholds as before. 
For model 1,
the strong lensing criterion is fulfilled for
38\% and 62\% of equivalent CG samples 5 and 10 times
more distant, respectively,  using the measured velocity dispersions for
$\sigma_v$,
and 69\% and 95\%, respectively, using the canonical value $\sigma_v = 315\,
\rm km \, s^{-1}$.
Thus, we obtain more efficient lensing in comparison with
MG94, who, with the same profile, predict
that 37\% of a sample of HCGs 10 times more distant will produce strong
lensing.
For our best fit model 2, we expect strong lensing for 18\% and 38\% of our
sample, respectively (see Fig. 2), using the measured velocity dispersions,
and 21\% and 64\% with $\sigma_v = 315 \, \rm km \, s^{-1}$.
These results confirm MG94's suggestion
that CGs in deeper surveys (and/or with larger velocity dispersion
than for HCGs)
could turn out to be efficient
strong lenses.

Lens systems cause an enhancement of the apparent magnitude of a source placed 
behind them.
Assuming a perfect alignment and $D_s = 2\,D_{ls}$,
the fraction of HCGs that amplify background sources by 
$\geq 0.5\,\rm mag$
is 21\% for model 1 and 8\% for models 2 and 3, using
measured $\sigma_v$s, and 26\% for model 1 and negligible for the others,
using $\sigma_v = 315 \, \rm km \, s^{-1}$. 
In comparison, MG94 find 16\%.

\section{DISCUSSION}

The uniqueness of the absolute surface density profiles of HCGs may seem
puzzling when considered within the hierarchy of 
galaxy systems. 
It arises from the characteristic mass density and velocity dispersion
required for virialized groups from a combination of sharply decreasing
cosmological distributions of these properties and minimum cutoffs for
virialization (Mamon 1994).
Consequently, virialized groups
have a characteristic radius, and by extension, a characteristic
density profile.

Because our method for determining the density profiles of groups does not
make use of a choice for a group center, only relying on the projected
separations, 
our fitted surface density profiles are  more concentrated than
that obtained by MG94.
Our profiles are even more concentrated than for most HCG diffuse
intergalactic gas 
distributions, as determined from X-rays.
Part of this inferred concentration may be due to small-scale correlations,
{\it i.e.\/,} binarity, in the galaxy distribution
(see Walke $\&$ Mamon 1989).
Note that, whereas Hickson (1982) only selected the bright galaxies, the
spectroscopically confirmed faint members of HCGs have somewhat larger
projected separations (Zepf, private communication).

Although the lensing efficiency
of HCGs is highly model dependent, we confirm MG94's result that HCGs should
not constitute strong lenses, as confirmed by the absence of giant arcs in
CCD images of HCGs (see Hickson 1993). 
CGs in 5 to 10 times deeper surveys should
produce a few giant arcs.
Our results do not exclude that a fraction of HCGs are due to projection
effects from loose groups or long filaments. In this case, real groups in
Hickson's catalog should be even more concentrated, and our conclusions on
lensing would be reinforced.

\acknowledgements

We thank the referee, Steve Zepf, and Yannick Mellier, for useful
comments.
This work was partially supported through
the Human Capital and Mobility Program
  of the EU (ULYSSES network) and  the Direcci\'on General
de Investigaci\'on Cient\'{\i}fica y
T\'ecnica (Spain)
(projects number PB93-0252,  AEN93--0673).
MLM thanks also the IAP (France),
where part of this work was done, for its kind hospitality.

\clearpage

\begin{deluxetable}{lccc}
\tablecaption{Kolmogorov-Smirnov test probabilities\label{tbl-1}}
\tablewidth{0pt}
\tablehead{
\colhead{} & \colhead{$s_{\rm min}$} & \colhead{$s_{\rm med}$} &
\colhead{$s_{\rm max}$}
}
\startdata
simulation-simulation & 0.987    & 0.909    & 0.752   \nl
simulation-catalog      &0.752    & 0.256   &  0.909   \nl
\enddata
\end{deluxetable}

\clearpage

\begin{figure}
\plotfiddle{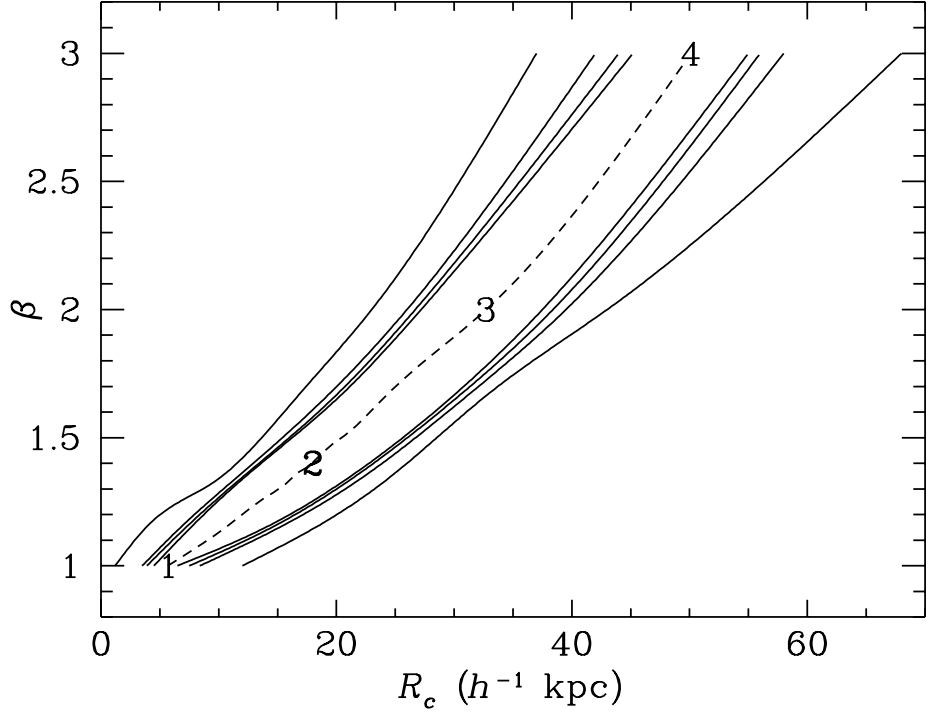}{0.2\vsize}{-90}{50}{50}{-200}{300}
\caption{Best-fit surface density profile (eq. [3]) parameters
of HCGs,
for confidence levels of $100\%$, $95\%$, $90\%$ and $85\%$ ({\it solid
curves}) and curve of optimum fits ({\it dashed curve}). Numbers refer
to our adopted models (with `2' our best solution).}
\label{fig1}
\end{figure}

\clearpage

\begin{figure}
\plotfiddle{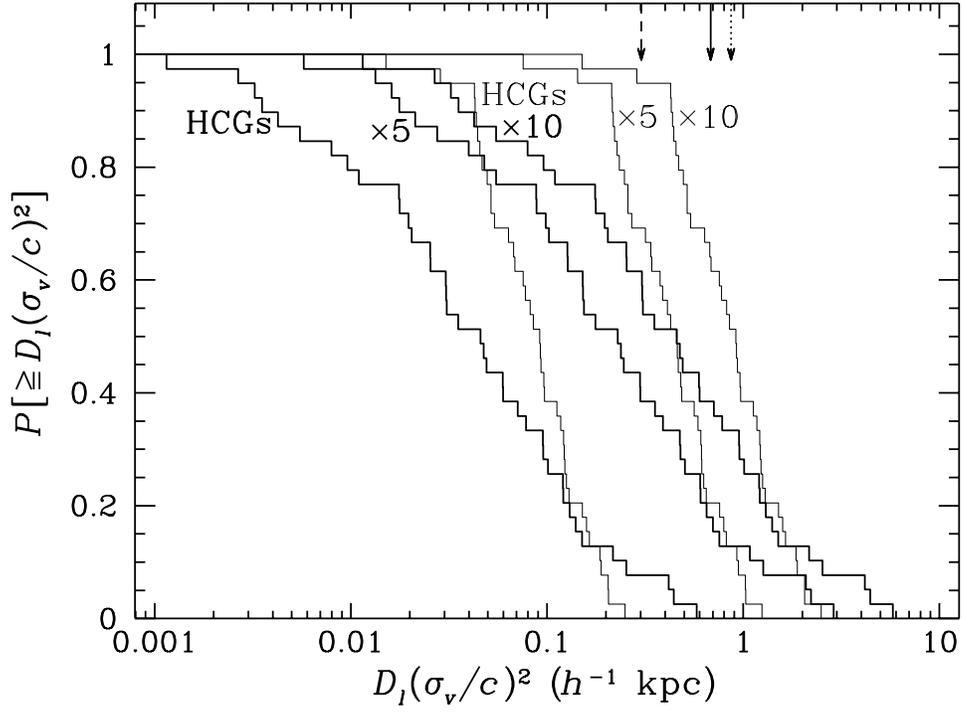}{0.2\vsize}{-90}{50}{50}{-200}{300}
\caption{Cumulative fraction of HCGs with values 
of $D_l (\sigma_v/c)^2$ greater than
given thresholds (eq. [6]), for our sample of HCGs, and for
the same samples placed at 
5 and 10 times their original distances.
Arrows show the thresholds for strong lensing satisfied for models 
 1 ({\it dashed arrow}), 2 ({\it solid
arrow\/}), and 3 
({\it dotted arrow}).
The threshold for model 4 is similar to that for model 3.
{\it Thick histograms} use measured velocity dispersions, while {\it thin
histograms}
use $\sigma_v = 315 \, \rm km \, s^{-1}$.}
\label{fig2}
\end{figure}

\end{document}